\documentclass[%
 amsmath,amssymb,
 aps,
]{revtex4-2}
\usepackage{amsmath}
\usepackage{float}
\usepackage{wrapfig}
\usepackage{graphicx}
\usepackage{dcolumn}
\usepackage[linesnumbered,ruled,vlined]{algorithm2e}
\usepackage{algpseudocode}
\usepackage{xcolor}
\usepackage{epstopdf}
\usepackage{appendix}
\usepackage{cases}
\usepackage{bm}

\begin{document}

\preprint{APS/123-QED}

\title{Rheology of three dimensional granular chute flows at large inertial numbers}

\author{Satyabrata Patro}
\affiliation{
 Department of Chemical Engineering, Indian Institute of Technology Kanpur, Uttar Pradesh, 208016, India
}
\author{Anurag Tripathi}%
\email{anuragt@iitk.ac.in}
\affiliation{
 Department of Chemical Engineering, Indian Institute of Technology Kanpur, Uttar Pradesh, 208016, India
}

\date{\today}

\begin{abstract}
The inertial number-based rheology, popularly known as the JFP model, is well known for describing the rheology of granular materials in the dense flow regime. While most of the recent studies focus on the steady-state rheology of granular materials, the time-dependent rheology of such materials has received less attention. Owing to this fact, we perform three-dimensional DEM simulations
of frictional inelastic spheres flowing down an inclined bumpy surface varying over a wide range of inclination angles and restitution coefficients. We show that steady, fully developed flows are possible at inclinations much higher than those predicted from the JFP model rheology. We show that, in addition to a modified effective friction law, the rheological description also needs to account for the stress anisotropy by means of a first and second normal stress difference law.
\end{abstract}

\maketitle
\section{\label{sec:level1}INTRODUCTION}

The rheology of granular materials has been an active research topic for the last few decades due to its wide occurrence in geophysical as well as industrial situations. \textcolor{black}{A number of experimental \cite{pouliquen1999scaling, midi2004dense, jop2005crucial, holyoake2012high, pouliquen1999shape, heyman2017experimental, thompson2007granular, lacaze2008planar,lube2004axisymmetric,lajeunesse2004spreading, zhou2017experiments, khakhar1997radial, annular_shear_cell_2009comparisons, azanza1999experimental, yang2001simulation,orpe2007rheology, arran2021laboratory, bachelet2023acoustic} as well as} simulation studies using discrete element method (DEM) \cite{silbert2001granular, midi2004dense, da2005rheophysics, pouliquen2006flow, baran_2006, borzsonyi2009patterns, tripathi2011rheology, kumaran2013, brodu_delannay_valance_richard_2015, mandal2016study, ralaiarisoa2017high, mandal2018study, bhateja2018, bhateja2020, patro2021rheology, debnath2022different,satya_time_dependent} have been utilized to explore the rheology of granular materials. A detailed review of granular flow rheology in different configurations can be found in \cite{forterre2008flows, andereotti_forterre_pouliquen_2013, jop2015rheological}. These studies have shown that the granular flow between the two limiting cases of quasistatic, slow flows, and rapid, dilute flows is controlled by a non-dimensional \textcolor{black}{inertial number $I=d\dot{\gamma}/\sqrt{P/\rho_p}$}, which depends on the local shear rate $\dot{\gamma}$ and pressure $P$ in addition to the particle size $d$ and density $\rho_p$. In our previous study, we investigated the rheology of two-dimensional granular chute flows at high inertial numbers and showed that the popularly used JFP model with a saturating behavior of the effective friction coefficient at high inertial numbers fails to capture the rheology at high inertial numbers \cite{patro2021rheology}. Instead, a non-monotonic variation of $\mu(I)$ is observed at large inertial numbers with a maximum at $I\approx 0.8$. The non-monotonic variation of the effective friction coefficient $\mu$, along with a weak power law relation of the solids fraction $\phi$ and a normal stress difference law $N_1/P$  with the inertial number $I$ has also been proposed in Patro \textit{et al.}\,\cite{patro2021rheology}  which describes the rheology at large inertial numbers.
Now, we explore the rheology of three-dimensional granular chute flows for a wide range of inclination angles and check whether the flow behavior observed in the 2D DEM simulations is observable in 3D DEM simulations as well.

Hence, we investigate the rheology of slightly polydisperse, inelastic spheres flowing down a rough inclined plane using discrete element method (DEM) simulations over a wide range of inclination angles. \textcolor{black}{Two different layer heights in the settled configuration of $L_y=20d$ and $L_y=40d$ are considered.} We show that steady, fully developed flows are possible at high inertial numbers $I>0.8$. We also show that in addition to the first normal stress difference which was found to be significant in $2D$, the second normal stress difference also becomes important in the case of $3D$ flows. \textcolor{black}{The restitution coefficient considered in this study is $e_n=0.5$. Due to the classical results of Silbert \textit{et al.}\,\cite{silbert2001granular}, which showed that the restitution coefficient has no significant role in the case of sufficiently frictional grains, most studies typically use normal coefficient of restitution close to $e_n=0.9$. Our results shown in \citet{patro2021rheology} show that using this default choice of restitution coefficient, the observation of transition from dense to dilute regime becomes very difficult since a small change in the inclination angle leads to accelerating flows. In order to obtain a large range of inclination angles at which non-accelerating flows are observable, we use these lower values of normal coefficient of restitution. We wish to emphasize the fact that the choice of restitution coefficient $e_n=0.5$ is not unrealistic and quite a few studies \cite{TRIPATHI2021288,lu_2019_bf,phdthesis_mitra,BARRIOS201384,WEI2020593,yu_en_value,yu_en_2,tang2021experimental,ye2019experimental} report the restitution coefficient of industrial grains near 0.5.} 

\section{\label{sec:level1}SIMULATION METHODOLOGY}

In the previous study, we investigated the rheology of two-dimensional granular chute flows at high inertial numbers and showed that the popularly used JFP model with a saturating behavior of the effective friction coefficient at high inertial numbers fails to capture the rheology at high inertial numbers. Instead, a non-monotonic variation of $\mu(I)$ is observed at large inertial numbers with a maximum at $I\approx 0.8$. The non-monotonic variation of the effective friction coefficient $\mu$, along with a weak power law relation of the solids fraction $\phi$ and a normal stress difference law $N_1/P$  with the inertial number $I$ has also been proposed in \citet{patro2021rheology} which describes the rheology at large inertial numbers.
Now, we explore the rheology of three-dimensional granular chute flows for a wide range of inclination angles and check whether the flow behavior observed in the 2D DEM simulations is observable in 3D DEM simulations as well.

Hence, we investigate the rheology of slightly polydisperse, inelastic spheres flowing down a rough inclined plane using discrete element method (DEM) simulations over a wide range of inclination angles. \textcolor{black}{Two different layer heights in the settled configuration of $L_y=20d$ and $L_y=40d$ are considered.} We show that steady, fully developed flows are possible at high inertial numbers $I>0.8$. We also show that in addition to the first normal stress difference which was found to be significant in $2D$, the second normal stress difference also becomes important in the case of $3D$ flows. \textcolor{black}{The restitution coefficient considered in this study is $e_n=0.5$. Due to the classical results of Silbert \textit{et al.}\,\cite{silbert2001granular}, which showed that the restitution coefficient has no significant role in the case of sufficiently frictional grains, most studies typically use normal coefficient of restitution close to $e_n=0.9$. Our results in \citet{patro2021rheology} show that using this default choice of restitution coefficient, the observation of transition from dense to dilute regime becomes very difficult since a small change in the inclination angle leads to accelerating flows. In order to obtain a large range of inclination angles at which non-accelerating flows are observable, we use these lower values of normal coefficient of restitution. We wish to emphasize the fact that the choice of restitution coefficient 0.5 is not unrealistic and quite a few studies \cite{TRIPATHI2021288,lu_2019_bf,phdthesis_mitra,BARRIOS201384,WEI2020593,yu_en_value,yu_en_2,tang2021experimental,ye2019experimental} report the restitution coefficient of industrial grains near 0.5.} 

\section{\label{sec:simulation_methodology}Simulation methodology}
\begin{figure}[hbtp]
    \centering
	\includegraphics[scale=0.38]{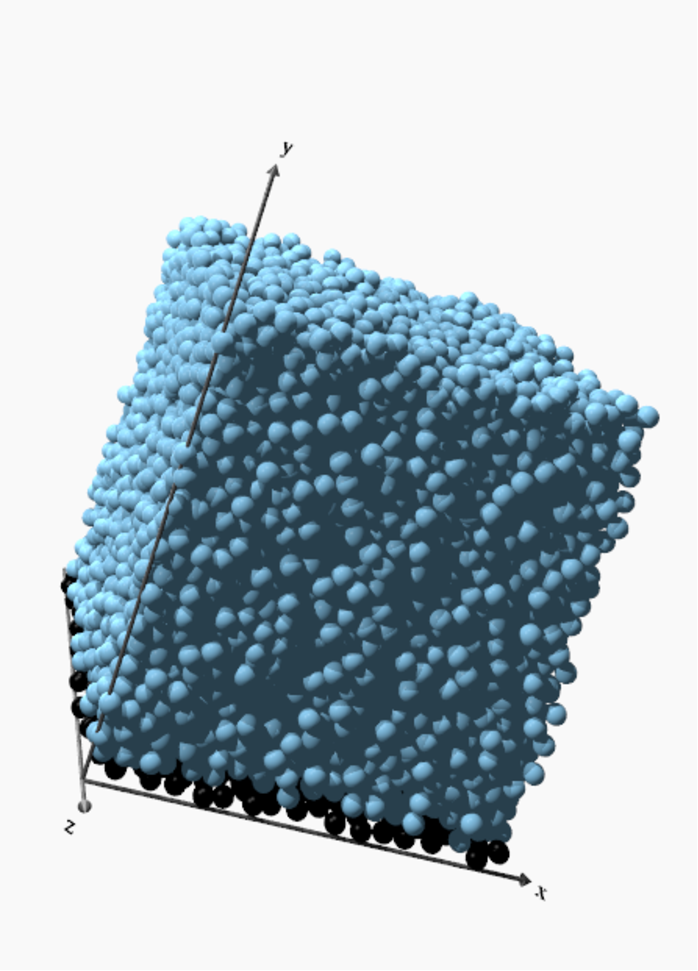}\put(-150,230){(a)}
 \hspace{2cm}
 \includegraphics[scale=0.38]{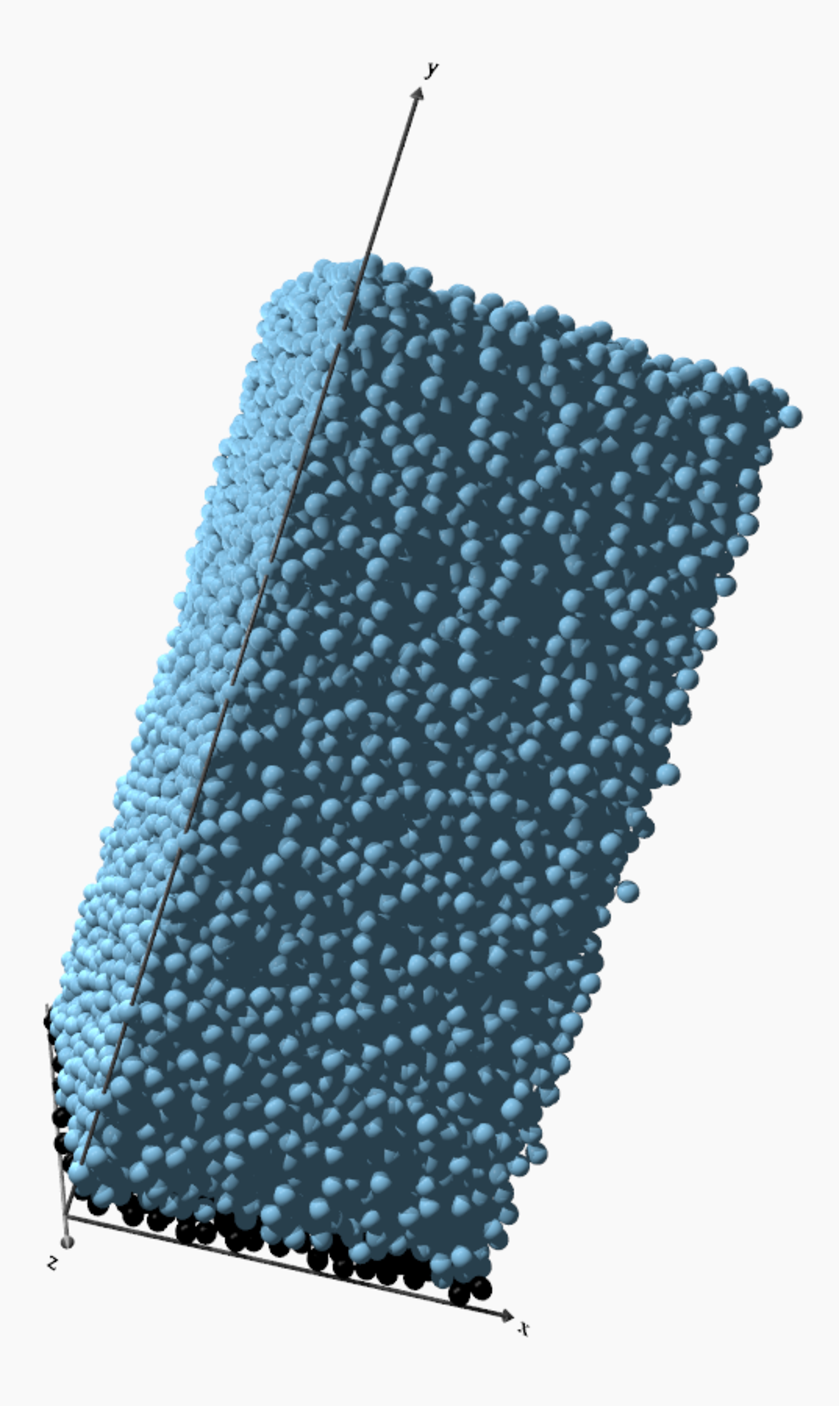}\put(-150,230){(b)}
	\caption{Chute flow simulation setup of dimension a) $20d\times20d\times20d$, and b) $20d\times40d\times20d$. black spheres represent the moving particles whereas the black spheres represent the static particles used to create a bumpy base. Periodic boundary conditions are used in the flow ($x$) and vorticity ($z$) direction.}
\label{Figure_1}
\end{figure}
The discrete element method (DEM) is used to perform three-dimensional simulations of slightly polydisperse particles ($\pm 5 \%$ polydispersity) of mean diameter $d=1mm$ flowing down a rough and bumpy inclined surface. The bumpy base is made of static spheres of diameter $2d$. The schematic view of the simulation setup is shown in Fig.~\ref{Figure_1} where the black spheres represent the flowing particles and black spheres represent the static particles of the bumpy base. The length and width of the simulation box is $L_x=L_z=20d$ along the $x$ and $z$ direction. The height of the simulation box is kept sufficiently large so that the particles do not feel the presence of the upper surface. In order to simulate an infinitely long inclined flow without end effects and neglect the variation in the vorticity direction, periodic boundary conditions are used in the flow ($x$) and vorticity ($z$) direction. Fig.~\ref{Figure_1}(a) shows the simulation setup where the settled layer height is $L_y=20d$ and fig.~\ref{Figure_1}(b) shows the simulation setup where the settled layer height is $L_y=40d$. The number of particles in the first case is $N_p=8000$ whereas the number of particles in the second case is $N_p=16000$.  
 
Particles are modeled as soft, deformable, inelastic, frictional spheres. The contact force between the spheres is modeled using the Hertz-Mindlin model in the LIGGGHTS-PUBLIC 3.0 open-source DEM package. The details about the contact force models, particle generation, particle settling, and property calculation are given in detail in \citet{patro2021rheology}.
\begin{figure}
\centering
\includegraphics[scale=0.6]{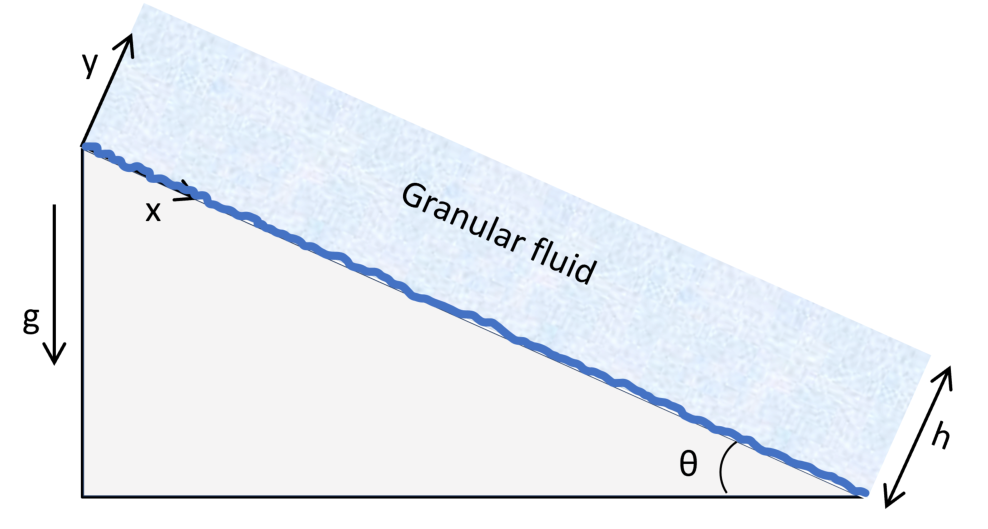}
\caption{Schematic of the chute flow geometry. The black-shaded region represents the granular fluid. The uneven surface near the base of the granular fluid represents the bumpy base.}
\label{Figure_2}
\end{figure}


Consider the steady and fully developed granular flow over a surface inclined at an angle $\theta$ under the influence of gravity. The schematic of granular fluid flowing down an inclined plane is shown in Figure \ref{Figure_2}.  Assuming a unidirectional flow in the x-direction, the
momentum balance equations in $x$ and $y$ direction simplify to
\begin{equation}
0=-\frac{\partial\tau_{yx}}{\partial{y}}+\rho{g}{\sin\theta},
\label{tau_yx_momentum_balance_x_1st_3d}
\end{equation}
\begin{equation}
0=-\frac{\partial{\sigma_{yy}}}{\partial{y}}-\rho{g}{\cos\theta}.
\label{sigma_yy_momentum_balance_y_1st_3d}
\end{equation}
Using the boundary conditions of zero shear stress and pressure at the free surface, \textcolor{black}{and assuming a constant density $\rho$ (i.e., constant solids fraction $\phi$) across the layer, eqs.~(\ref{tau_yx_momentum_balance_x_1st_3d}) and (\ref{sigma_yy_momentum_balance_y_1st_3d}) can be integrated to obtain}
\begin{equation}
\tau_{yx}=-\rho g(h-y)\sin\theta,
\label{tau_yx_momentum_balance_x_3d}
\end{equation}
\begin{equation}
\sigma_{yy}=\rho g(h-y)\cos\theta.
\label{sigma_yy_momentum_balance_y_3d}
\end{equation}
The effective friction coefficient $\mu(I)$, defined as the ratio of the magnitude of shear stress $|\tau_{yx}|$ to the confining pressure $P$, i.e., 
\begin{equation}
\mu(I)=\frac{|\tau_{yx}|}{P}
\label{mu_I_general_equation_3d}
\end{equation} 
is dependent only on the dimensionless inertial number $I$ in the dense flow regime.
Mandal \textit{et al.}\,\cite{mandal2016study} have proposed a non-monotonic variation of the effective friction coefficient $\mu(I)$ as
\begin{equation}
\mu(I)=\mu_{s}+\frac{c_{1}-c_{2}I}{1+I_{0}/I}~,
\label{mu_I_modified_JFP_3d}
\end{equation}
where $\mu_s$, $I_{0}$, $c_1$ and $c_2$ are the model parameters. The solids fraction $\phi$ shows a power law variation with the inertial number $I$ and is given as
\begin{equation}
\phi(I)=\phi_{max}-aI^{\alpha},
\label{phi_power_law_model_3d}
\end{equation}
where $\phi_{max}$, $a$ and $\alpha$ are the model parameters.
Simulation results in this study show that the first normal stress difference $N_1=\sigma_{xx}-\sigma_{yy}$ to the pressure $P$ ratio as well as the second normal stress difference $N_2=\sigma_{yy}-\sigma_{zz}$ to the pressure $P$ ratio are a function of inertial number $I$, i.e.,
$\frac{N_{1}}{P}=f_1(I)$, and 
$\frac{N_{2}}{P}=f_2(I)$
where $P=(\sigma_{xx}+\sigma_{yy}+\sigma_{zz})/3$ is the trace of the stress tensor.
These functional forms can be used to relate the pressure $P$ with the first as well as second normal stress difference $\sigma_{yy}$ as
\begin{equation}
P=\frac{\sigma_{yy}}{1+\dfrac{f_2(I)-f_1(I)}{3}}.
\label{P_sigmayy_3d}
\end{equation}
Using Eq.~(\ref{sigma_yy_momentum_balance_y_3d}) we obtain,
\begin{equation}
P=\frac{\rho g(h-y)\cos\theta}{1+\dfrac{f_2(I)-f_1(I)}{3}}.
\label{newPressure_3d}
\end{equation}
Using the expression for inertial number $I=\frac{d\dot\gamma}{\sqrt{P/\rho_p}}$ with $\dot\gamma=\frac{dv_x}{dy}$ and rearranging, we get
\begin{equation}
\frac{d{v_x}}{d{y}}=\frac{I\sqrt{P/\rho_p}}{d}.
\label{solve_for_v_3d}
\end{equation}
Using Eq.~(\ref{newPressure_3d}), Eq.~(\ref{solve_for_v_3d}) can be integrated to obtain the velocity in the $x$-direction as
\begin{equation}
v_x=v_{slip}+\frac{2}{3}\frac{I}{d}\sqrt{\frac{\phi g\cos\theta}{1+\dfrac{f_2(I)-f_1(I)}{3}}}\left[h^{3/2}-\left(h-y\right)^{3/2}\right],
\label{vx}
\end{equation}
where $v_{slip}$ is the slip velocity at the base.
Using Eqs.~(\ref{tau_yx_momentum_balance_x_3d}) and (\ref{newPressure_3d}) in Eq.~(\ref{mu_I_general_equation_3d}), the effective friction coefficient $\mu$ at any inclination $\theta$ is obtained as
\begin{equation}
\mu(I)=\dfrac{\tan\theta}{1+\dfrac{f_2(I)-f_1(I)}{3}}.
\label{new_muI_3d}
\end{equation}
Using our simulation data over a large range of $I$, we find that $f_1(I)$ shows a linear variation for the range of the inertial number considered in this study. We also find that $f_2(I)$ shows a quadratic variation with the inertial number.
Thus we can express $f_1(I)=AI+B$ and $f_2(I)=CI^2+DI+E$ for the range of inertial number of our interest. Using Eq.~(\ref{mu_I_modified_JFP_3d}) along with the linear form of $f_1(I)$ and quadratic form of $f_2(I)$ in Eq.~(\ref{new_muI_3d}), we obtain  an algebraic equation $G(I)=A_0I^3+B_0I^2+C_0I+D_0=0$. 

This algebraic equation needs to be solved to obtain the value of the inertial number at a given inclination $\theta$. 

By using $f_1(I)=AI+B$ and $f_2(I)=CI^2+DI+E$, we get the coefficients of the cubic equation as $A_0=-C\tan\theta$, $B_0=(A-D-CI_0)\tan\theta-c_2$, $C_0=(-3+B-E+AI_0-DI_0)\tan\theta+3\mu_s+3\mu_m$ and $D_0=((-3+B-E)\tan\theta+3\mu_s)I_0$. 

Solving the cubic equation $G(I)=0$, we obtain three different possible values of inertial number $I$ for a particular inclination $\theta$. Only one of these values is found to be realistic. Using this value of the inertial number $I$ for a given inclination angle $\theta$, the solids fraction $\phi$ is obtained using Eq.~(\ref{phi_power_law_model_3d}).
With the increase in the inclination angle $\theta$, the inertial number $I$ increases, and solids fraction $\phi$ decreases leading to an increase in the overall height of the flowing layer. This increased flowing layer height $H$ at any inclination $\theta$ can be calculated from the mass balance over the entire layer using the following equation
\begin{equation}
H = H_{min}\phi_{max}/\phi,
\label{Hphi_3d}
\end{equation}
\textcolor{black}{where $H_{min}$ is the minimum height of the static layer having maximum solids fraction $\phi_{max}$. This value of $H$ is used to get the velocity $v_x(y)$ using Eq. \ref{vx}}

\section{\label{sec:Results_and_discussions}Results and Discussion}
We present DEM simulation results for the flow of spheres over a bumpy inclined surface starting from two different settled layer heights of $L_y=20d$ and $L_y=40d$ for restitution coefficient $e_n=0.5$ where $d$ is the mean particle diameter. 
All the quantities reported are non-dimensionalized using $d$, $m$, $\sqrt{d/g}$ and $mg$ as length, mass, time, and force units respectively where $m$ is the mass of the particle having diameter $d$ and $g$ is the gravitational acceleration. 

\subsection{Existence of steady flows}

\begin{figure*}[hbtp]
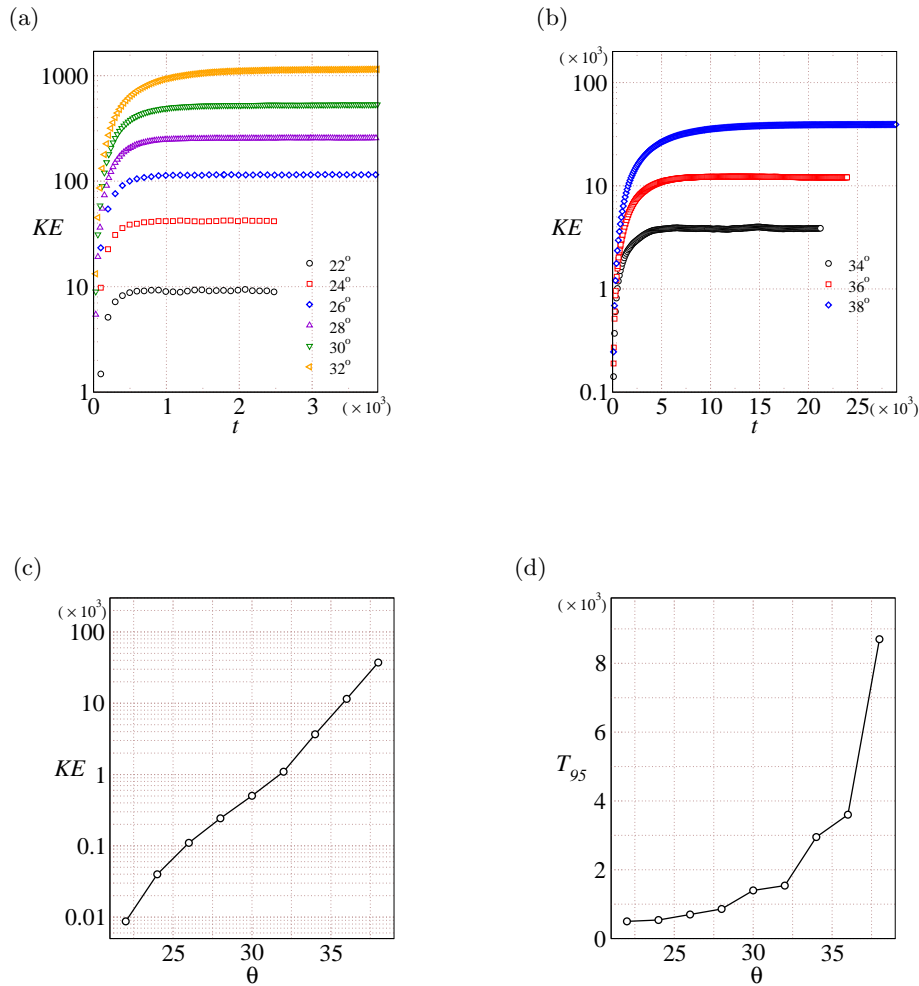

\centering
	\includegraphics[scale=0.35]{Fig_3a.eps}
 	\put (-145,155) {(a)}
 	\hspace{2cm}
	\includegraphics[scale=0.35]{Fig_3b.eps}
 	\put (-145,155) {(b)}

\vspace{1.6cm}
	\includegraphics[scale=0.35]{Fig_3c.eps}
	\put (-145,155){(c)}
	\hspace{2cm}
	\includegraphics[scale=0.35]{Fig_3d.eps}
	\put (-145,155){(d)}
 	\caption {Variation of the average kinetic energy $KE$ with time $t$ for (a) inclination angles $\theta\leq32^\circ$ and (b) inclination angles $\theta>32^\circ$. Variation of (c) steady-state kinetic energy and (d) time taken to reach 95\% of steady-state kinetic energy with the inclination angle for $e_n=0.5$.}
  	\label{Figure_3}
\end{figure*}

We perform DEM simulations for inelastic spheres having restitution coefficient $e_n=0.5$ spanning a wide range of inclination angles $\theta=22^\circ$ to $38^\circ$ for the initial flowing layer thickness $L_y=40d$. The average kinetic energy of the particles at these inclinations are shown in Fig.~\ref{Figure_3}(a) and Fig.~\ref{Figure_3}(b). The average value of the kinetic energy keeps increasing with time and eventually becomes constant, indicating that the system is able to achieve a steady state at these inclinations.
The average kinetic energy of the particles, shown in Fig.~\ref{Figure_3}(a) and Fig.~\ref{Figure_3}(b), becomes constant for all the inclination angles considered. This confirms that steady, fully developed flows are possible over a larger range of inclinations for $e_n=0.5$. 
The average kinetic energy of the particles at steady state is plotted with the inclination angle in Fig.~\ref{Figure_3}(c). As expected, the kinetic energy increases with inclination. A linear increase in the average kinetic energy is observed with the inclination angle $\theta$. Fig.~\ref{Figure_3}(d) shows the time taken to achieve $95\%$ of steady state kinetic energy, denoted as $T_{95}$, for different inclinations. The time taken to reach a steady state increases with the inclination angle and a sharp change is observable after $\theta=32^\circ$. Note that the kinetic energy in the $y$ axis is non-dimensionalized using $mgd$ where $m$ is the mass, $d$ is the particle diameter and $g$ is the acceleration due to gravity.

\begin{figure*}[hbtp]
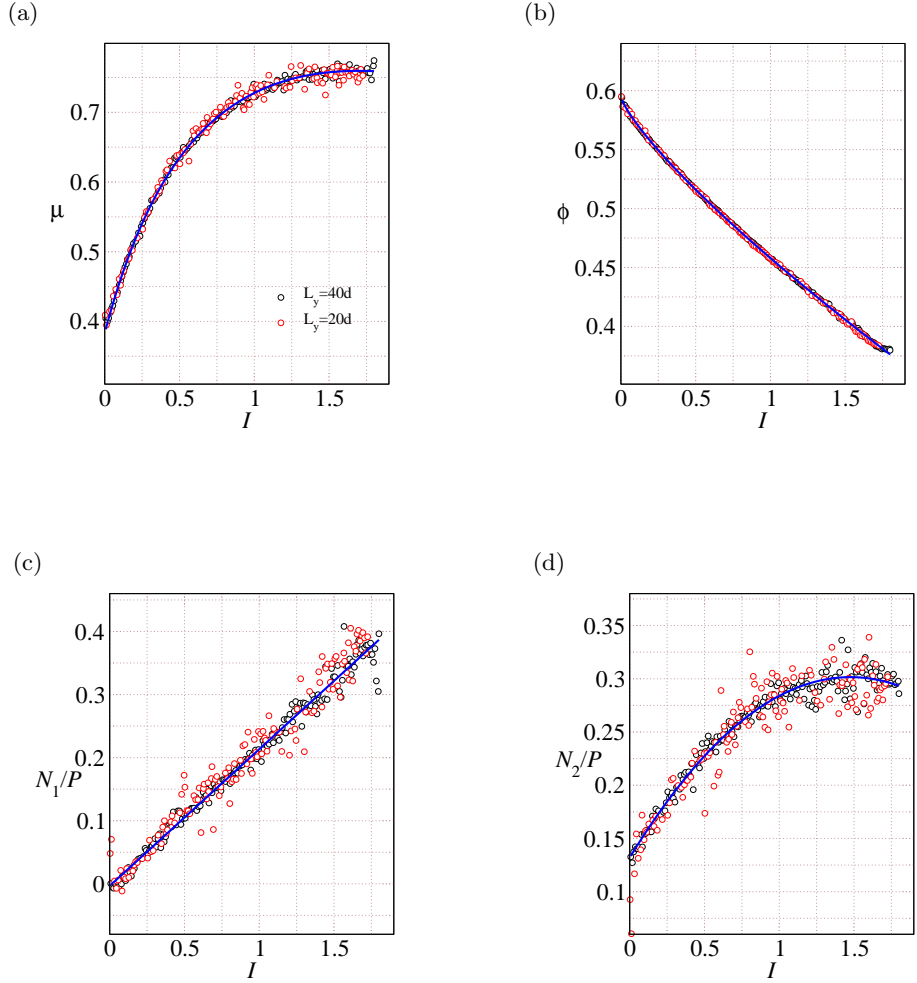

	\centering
		\includegraphics[scale=0.35]{Fig_4a.eps}
		\put (-145,155){(a)}
		\hspace{2cm}
		\includegraphics[scale=0.35]{Fig_4b.eps}
		\put (-145,155){(b)}

  \vspace{1.6cm}
 		\includegraphics[scale=0.35]{Fig_4c.eps}
		\put (-145,155){(c)}
        \hspace{2cm}
        \includegraphics[scale=0.35]{Fig_4d.eps}
		\put (-145,155){(d)}
	\caption {Variation of the effective friction coefficient $\mu$, solids fraction $\phi$, first normal stress difference to the pressure ratio $N_1/P$, and second normal stress difference to the pressure ratio $N_2/P$ with the inertial number $I$ for $e_n=0.5$. Symbols represent simulation data and lines represent best-fit curves. (a) Effective friction coefficient with best fit using Eq.~(\ref{mu_I_modified_JFP_3d}) (solid line). 
	(b) Solids fraction with best fit using Eq.~(\ref{phi_power_law_model_3d}). (c) First normal stress difference to the pressure ratio (symbols), with the best fit (line). Data for the entire range of $I$ are fitted using $f_1(I)=AI+B$. (d) Second normal stress difference to the pressure ratio (symbols), with the best fit (line). Data for the entire range of $I$ are fitted using $f_2(I)=CI^2+DI+E$. Black circles correspond to the DEM simulations starting from a settled layer height of $L_y=40d$ and red circles correspond to the DEM simulations starting from a settled layer height of $L_y=20d$. The black solid lines represent the best fit to the DEM data using appropriate equations (See text for details).}
	\label{Figure_4}
\end{figure*}

\subsection{Rheological model}

From the DEM simulations data obtained from two different flowing layer thicknesses $L_y=40d$ and $L_y=20d$, we report the variation of the effective friction coefficient $\mu$, solids fraction $\phi$, the ratio of first normal stress difference to pressure $N_1/P$ and the ratio of second normal stress difference to pressure $N_2/P$ with inertial number $I$ for $e_n=0.5$ in Fig.~\ref{Figure_4}. The black circles correspond to the DEM simulations starting from a settled layer height of $L_y=40d$ and the red circles correspond to the DEM simulations starting from a settled layer height of $L_y=20d$. \textcolor{black}{The data for the two cases do not differ significantly from each other and can be described using a single curve.} The black solid line represents the best fit to the data accounting for both layer heights.  
The effective friction coefficient obtained from the simulations (shown using black and red symbols in Fig.~\ref{Figure_4}(a)) increases up to $I\approx1.8$ and shows a saturating behavior at large inertial numbers. There is a slight decrease in the effective friction coefficient $\mu$ at large inertial numbers $I$, however, the decrease is not prominent, and the inclusion of $\mu-I$ data for inclination angles $\theta>38^\circ$ is needed to confirm the decrease in the effective friction coefficient at large inertial numbers. \textcolor{black}{Given the sharp increase in the computational time for higher inclinations, we proceed with the analysis of the existing data.}

Fitting the 4-parameter MK model using Eq.~(\ref{mu_I_modified_JFP_3d}) to the simulation data shown in Fig.~\ref{Figure_4}(a) indicates that the MK model (shown using solid lines) is able to describe the $\mu-I$ variation very well. Note that the simulation data for the inclination angle $\theta=22^\circ-38^\circ$ is considered to obtain the model parameters. The MK model parameters are reported in Table \ref{parameter_mk_model_3d_40d}. 
The variation of solids fraction is plotted with the inertial number in Fig.~\ref{Figure_4}(b). The solid line shown in Fig.~\ref{Figure_4}(b) is obtained by fitting the power law variation of $\phi$ with $I$ (Eq.~(\ref{phi_power_law_model_3d})) to the simulation data. The dilatancy law model parameters are reported in Table \ref{parameters_dilatancy_power_law_3d_40d}. 
In  Fig.~\ref{Figure_4}(c), we report the variation of the ratio of the first normal stress difference to the pressure ratio $N_1/P$ with the inertial number $I$. The variation for the first normal stress difference to the pressure ratio $N_1/P$ with the inertial number $I$ is described well by a linear relation of the form $N_1/P=f_1(I)=AI+B$. The first normal stress difference law model parameters are reported in Table \ref{parameter_n1_by_p_3d_40d}. We also report the variation of the ratio of the second normal stress difference to the pressure ratio $N_2/P$ with the inertial number $I$. The variation for the second normal stress difference to the pressure ratio $N_2/P$ with the inertial number $I$ is described well by a quadratic relation of the form $N_2/P=f_2(I)=FI^2+GI+H$. The second normal stress difference law model parameters are reported in Table \ref{parameter_n2_by_p_3d_40d}. \textcolor{black}{A comparison of Fig.~\ref{Figure_4}(c) and Fig.~\ref{Figure_4}(d) shows that while $N_1/P$ is zero at low inertial numbers, $N_2/P$ is found to be positive even for the quasistatic systems with $I\sim0$. In addition, $N_2/P>N_1/P$ for $I\leq1.25$. For $I>1.25$, however, $N_1/P$ becomes larger than $N_2/P$. This transition of $N_2/P-N_1/P$ from positive to negative at $I\sim1.25$ needs to be explored further.}
\begin{table}[h]
\caption{\label{parameter_mk_model_3d_40d}Model parameters for MK model}
\centering
\begin{tabular}{ccccc}
$e_n$ & $\mu_s$	& $c_1$ & $c_2$	& $I_0'$\\  
\hline
0.5  & 0.383 & 0.821 & 0.131 & 1.002\\ 
\end{tabular}
\end{table}

\begin{table}[h]
\caption{\label{parameters_dilatancy_power_law_3d_40d}Model parameters for dilatancy power law}
\centering
\begin{tabular}{cccc}
$e_n$ & $\phi_{max}$ & $a$ & $\alpha$ \\ \hline
0.5 & 0.595 & 0.137 & 0.797 \\
\end{tabular}
\end{table}

\begin{table}[h]
\centering
\caption{\label{parameter_n1_by_p_3d_40d}Model parameters for linear $N_1/P=AI+B$.}  
\begin{tabular}{ccc}
$e_n$ & $A$ & $B$ \\ 
\hline
0.5 & 0.217 & -0.003 \\
\end{tabular}
\end{table}

\subsection{Theoretical predictions }

\begin{figure*}[h]
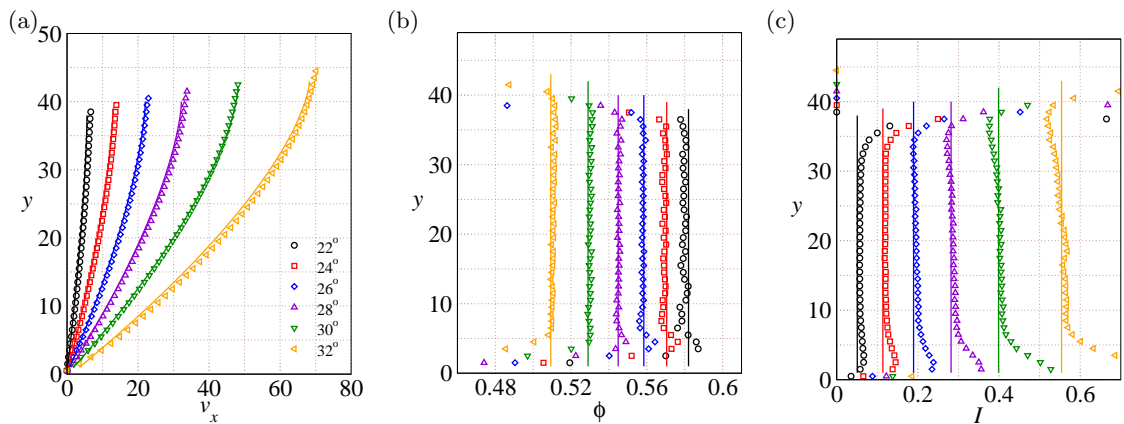

 	\centering
	
	\includegraphics[scale=0.35]{Fig_5a.eps}
 	\put (-135,150) {(a)}
	\hspace{0.5cm}
	\includegraphics[scale=0.35]{Fig_5b.eps}
    \put (-135,150){(b)}
	\hspace{0.5cm}
	\includegraphics[scale=0.35]{Fig_5c.eps}
	\put (-135,150){(c)}
	\caption {Variation of the (a) velocity $v_x$, (b) solids fraction $\phi$, and (c) inertial number $I$ across the layer $y$ at inclination angles $\theta=22^\circ-32^\circ$. All data correspond to $e_n=0.5$. Solid lines represent the theoretical predictions (see text for details). The height of the settled layer at $t=0$ is $40d$.}
		\label{Figure_5}
\end{figure*}

Using the rheological model parameters given in Table \ref{parameter_mk_model_3d_40d}-\ref{parameter_n2_by_p_3d_40d}, we predict various flow properties of interest. These theoretical predictions are reported as solid lines in Figs.~\ref{Figure_5} and \ref{Figure_6}. The velocity profiles predicted using the theory are found to be in excellent agreement with the simulation results for inclination angles ranging from $\theta=22^\circ-32^\circ$ (see Fig.~\ref{Figure_5}(a)). The slip velocity, which appears to be small for these inclinations is taken from the simulation data. The predicted values of the solids fraction $\phi$ are also found to be in very good agreement for inclination angles $\theta=22^\circ$ to $\theta=32^\circ$ as shown in Fig.~\ref{Figure_5}(b). The theoretical inertial numbers are also found to be in good agreement with the inertial numbers obtained from DEM simulations (see Fig.\ref{Figure_6}(c)).

\begin{table}[h]
\centering
\caption{\label{parameter_n2_by_p_3d_40d}Model parameters for quadratic $N_2/P=FI^2+GI+H$} 
\begin{tabular}{cccc}
$e_n$ & $F$ & $G$ & $H$ \\ \hline
0.5 & -0.078 & 0.23 & 0.132 \\
\end{tabular}
\end{table}

\begin{figure*}[hbp]
 	\centering
	\includegraphics[scale=0.35]{Fig_6a.eps}
	\put (-135,150){(a)}
	\hspace{2cm}
	\includegraphics[scale=0.35]{Fig_6b.eps}
	\put (-135,150){(b)}

 \vspace{1.6cm}
	\includegraphics[scale=0.35]{Fig_6c.eps}
	\put (-135,150){(c)}
 \hspace{2cm}
 \includegraphics[scale=0.35]{Fig_6d.eps}
	\put (-135,150){(d)}
		\caption {Variation of the (a) pressure $P$, (b) shear stress $\tau_{yx}$ (c) first normal stress difference $N_1$ and (d) second normal stress difference $N_2$ across the layer $y$ at inclination angles $\theta=22^\circ-32^\circ$. All data correspond to $e_n=0.5$. Solid lines represent the theoretical predictions (see text for details). The height of the settled layer at $t=0$ is $40d$.}
		\label{Figure_6}
	\end{figure*}

Figs.\ref{Figure_6}(a-d) shows the comparison between DEM simulations and the theoretical predictions for the pressure $P$, shear stress $\tau_{yx}$, first normal stress difference $N_1$, and the second normal stress difference $N_2$ for the range of inclinations $\theta=22^\circ$ to $\theta=32^\circ$. The theoretical predictions using the rheological parameters are in very good agreement with the DEM simulation results. \textcolor{black}{Comparison of $N_1$ and $N_2$ shows that $N_2$ is higher than $N_1$ for these range of inclinations.}

\begin{figure*}
 	\centering
	\includegraphics[scale=0.35]{Fig_7a.eps}
 	\put (-135,150) {(a)}
	\hspace{0.5cm}
	\includegraphics[scale=0.35]{Fig_7b.eps}
    \put (-135,150){(b)}
	\hspace{0.5cm}
	\includegraphics[scale=0.35]{Fig_7c.eps}
	\put (-135,150){(c)}
	
		\caption {Variation of the (a) velocity $v_x$, (b) solids fraction $\phi$, and (c) inertial number $I$ across the layer $y$ at inclination angles $\theta=34^\circ-38^\circ$. All data correspond to $e_n=0.5$. Solid lines represent the theoretical predictions (see text for details). The height of the settled layer at $t=0$ is $40d$.}
		\label{Figure_7}
	\end{figure*}
\vspace{1.6cm}
\begin{figure*}
 	\centering
	\includegraphics[scale=0.35]{Fig_8a.eps}
	\put (-135,150){(a)}
	\hspace{2cm}
	\includegraphics[scale=0.35]{Fig_8b.eps}
	\put (-135,150){(b)}

\vspace{1.6cm}
 
	\includegraphics[scale=0.35]{Fig_8c.eps}
	\put (-135,150){(c)}
 \hspace{2cm}
 \includegraphics[scale=0.35]{Fig_8d.eps}
	\put (-135,150){(d)}
		\caption {Variation of the (a) pressure $P$, (b) shear stress $\tau_{yx}$ (c) first normal stress difference $N_1$ and (d) second normal stress difference $N_2$ across the layer $y$ at inclination angles $\theta=34^\circ-38^\circ$. All data correspond to $e_n=0.5$. Solid lines represent the theoretical predictions (see text for details). The height of the settled layer at $t=0$ is $40d$.}
		\label{Figure_8}
	\end{figure*}

We next use the rheological model parameters given in Table \ref{parameter_mk_model_3d_40d}-\ref{parameter_n2_by_p_3d_40d} to predict various flow properties at large inclination angles ranging from $\theta=34^\circ-38^\circ$. These theoretical predictions are reported as solid lines in Figs.~\ref{Figure_7} and \ref{Figure_8}. The velocity profiles predicted using the theory are found to be in good agreement with the simulation results even at large inclination angles ranging from $\theta=34^\circ-38^\circ$ (see Fig.~\ref{Figure_7}(a)). We use the slip velocity obtained from DEM simulations in the theoretical predictions. The predicted values of the solids fraction $\phi$ are also found to be in very good agreement for inclination angles $\theta=34^\circ$ to $\theta=38^\circ$ as shown in Fig.~\ref{Figure_7}(b). In addition, the theoretical inertial numbers are also found to be in good agreement with the inertial numbers obtained from DEM simulations (see Fig.\ref{Figure_8}(c)) except the largest inclination angle $\theta=38^\circ$ where the theory predicts the inertial number $I=1.76$ in comparison to the inertial number $I=1.9$ obtained from DEM simulations.
Figs. \ref{Figure_8}(a)-(d) show the comparison between DEM simulations and the theoretical predictions for the pressure $P$, shear stress $\tau_{yx}$, first normal stress difference $N_1$, and the second normal stress difference $N_2$ for the large inclinations angles ranging from $\theta=34^\circ$ to $\theta=38^\circ$. The theoretical predictions using the rheological parameters are in very good agreement with the DEM simulation results.  

\begin{figure*}
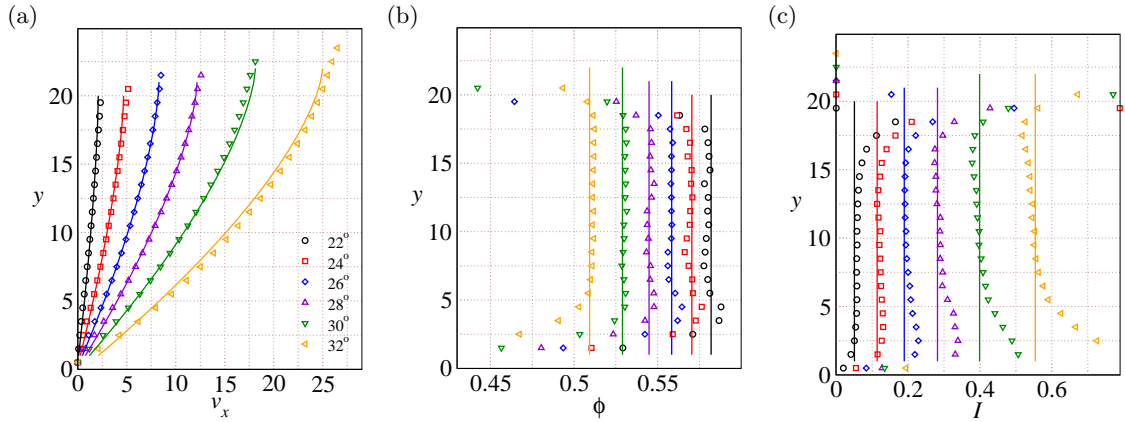

 	\centering
	\includegraphics[scale=0.35]{Fig_9a.eps}
 	\put (-135,150) {(a)}
	\hspace{0.5cm}
	\includegraphics[scale=0.35]{Fig_9b.eps}
    \put (-135,150){(b)}
	\hspace{0.5cm}
	\includegraphics[scale=0.35]{Fig_9c.eps}
	\put (-135,150){(c)}
	\caption {Variation of the (a) velocity $v_x$, (b) solids fraction $\phi$, and (c) inertial number $I$ across the layer $y$ for inclination angles ranging from $\theta=22^\circ$ to $\theta=32^\circ$ for $20d$ thickness of the layer. All data correspond to $e_n=0.5$. Solid lines represent the theoretical predictions using the model parameters obtained from DEM simulation data.}
		\label{Figure_9}
	\end{figure*}
\vspace{1.6cm}
\begin{figure*}
 	\centering
	\includegraphics[scale=0.35]{Fig_10a.eps}
	\put (-135,150){(a)}
	\hspace{2cm}
	\includegraphics[scale=0.35]{Fig_10b.eps}
	\put (-135,150){(b)}

\vspace{1.6cm}
 
	\includegraphics[scale=0.35]{Fig_10c.eps}
	\put (-135,150){(c)}
 \hspace{2cm}
 \includegraphics[scale=0.35]{Fig_10d.eps}
	\put (-135,150){(d)}
		\caption {Variation of the (a) pressure $P$, (b) shear stress $\tau_{yx}$ (c) first normal stress difference $N_1$ and (d) second normal stress difference $N_2$ across the layer $y$ for inclination angles ranging from $\theta=22^\circ$ to $\theta=32^\circ$. All data correspond to $e_n=0.5$. Solid lines represent the theoretical predictions using the model parameters obtained from DEM simulations data.}
		\label{Figure_10}
	\end{figure*}
 Many industrial granular flows in conveyors and chutes have flowing layer thickness in the range $10d-20d$. It is well known that the inertial number $I$ and solids fraction $\phi$ are nearly constant in the bulk region of the flowing layer of chute flow for any given inclination angle $\theta$. \textcolor{black}{The bulk region is defined as the region that is away from the base of the chute and the free surface. The bulk region is very important in order to estimate the rheological properties. Lack of bulk region in the case of low flowing layer thickness will lead to fluctuations of flow properties in the bulk and will lead to significant errors in estimating the rheological parameters. Due to the lack of bulk region in the case of low flowing layer thickness $L_y=20d$, we have performed DEM simulations starting from a settled flowing layer height $L_y=40d$.} We will now use the rheological model parameters given in Table \ref{parameter_mk_model_3d_40d}-\ref{parameter_n2_by_p_3d_40d} obtained from the flowing layer thickness $L_y=40d$ to predict various flow properties for a wider range of inclination angles ranging from $\theta=22^\circ-38^\circ$ for low flowing layer thickness $L_y=20d$ which is the initial settled flowing layer thickness.

\begin{figure*}
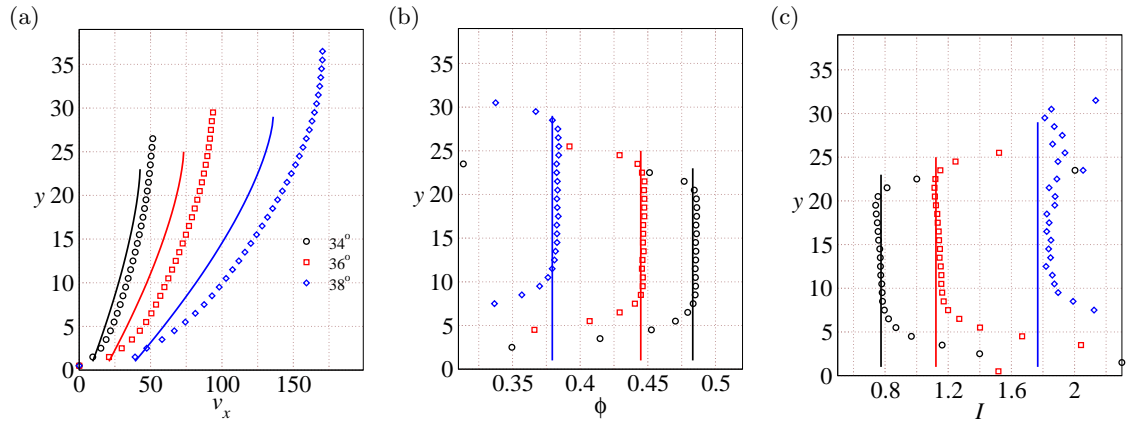

 	\centering
	\includegraphics[scale=0.35]{Fig_11a.eps}
 	\put (-135,150) {(a)}
	\hspace{0.5cm}
	\includegraphics[scale=0.35]{Fig_11b.eps}
    \put (-135,150){(b)}
	\hspace{0.5cm}
	\includegraphics[scale=0.35]{Fig_11c.eps}
	\put (-135,150){(c)}
	\caption {Variation of the (a) velocity $v_x$, (b) solids fraction $\phi$, and (c) inertial number $I$ across the layer $y$ for inclination angles ranging from $\theta=34^\circ$ to $\theta=38^\circ$. All data correspond to $e_n=0.5$. Solid lines represent the theoretical predictions using the model parameters obtained from DEM simulations data of $L_y=40d$ layer thickness.}
		\label{Figure_11}
	\end{figure*}
\vspace{1.6cm}
\begin{figure*}
 	\centering
	\includegraphics[scale=0.35]{Fig_12a.eps}
	\put (-135,150){(a)}
	\hspace{2cm}
	\includegraphics[scale=0.35]{Fig_12b.eps}
	\put (-135,150){(b)}

\vspace{1.6cm}
 
	\includegraphics[scale=0.35]{Fig_12c.eps}
	\put (-135,150){(c)}
 \hspace{2cm}
 \includegraphics[scale=0.35]{Fig_12d.eps}
	\put (-135,150){(d)}
		\caption {Variation of the (a) pressure $P$, (b) shear stress $\tau_{yx}$ (c) first normal stress difference $N_1$ and (d) second normal stress difference $N_2$ across the layer $y$ for inclination angles ranging from $\theta=34^\circ$ to $\theta=38^\circ$. All data correspond to $e_n=0.5$. Solid lines represent the theoretical predictions using the model parameters obtained from DEM simulations data of $L_y=40d$ layer thickness.}
\label{Figure_12}
	\end{figure*}
 
 These theoretical predictions are reported as solid lines in Figs.~\ref{Figure_9} and \ref{Figure_10}. The velocity profiles predicted using the theory are found to be in good agreement with the simulation results for inclination angles ranging from $\theta=22^\circ-32^\circ$ (see Fig.~\ref{Figure_9}(a)). Figs.~\ref{Figure_9}(b) and Fig.~\ref{Figure_9}(c) show that the predicted values of the solids fraction $\phi$ and inertial number $I$ are also found to be in very good agreement for inclination angles $\theta=22^\circ$ to $\theta=32^\circ$.
Figs.\ref{Figure_10}(a-d) shows the comparison between DEM simulations and the theoretical predictions for the pressure $P$, shear stress $\tau_{yx}$, first normal stress difference $N_1$, and the second normal stress difference $N_2$ for inclinations angles ranging from $\theta=22^\circ$ to $\theta=32^\circ$. Again, the theoretical predictions using the rheological parameters are in very good agreement with the DEM simulation results. Next, the theoretical predictions for the various flow properties of interest have been compared with the DEM simulation results for higher inclination angles in the range of $\theta=34^\circ$ to $\theta=38^\circ$. The predictions from the rheological parameters obtained using the simulations data of $L_y=40d$ layer are in very good agreement with the DEM simulation results for low flowing layer thickness $L_y=20d$ suggesting that rheological parameters obtained from large flowing layer thickness data can be used to predict the flow properties for low flowing layer thickness as well (shown in Figs.\ref{Figure_11}-\ref{Figure_12}).

\subsection{Bulk average properties}

\begin{figure*}[hbtp]
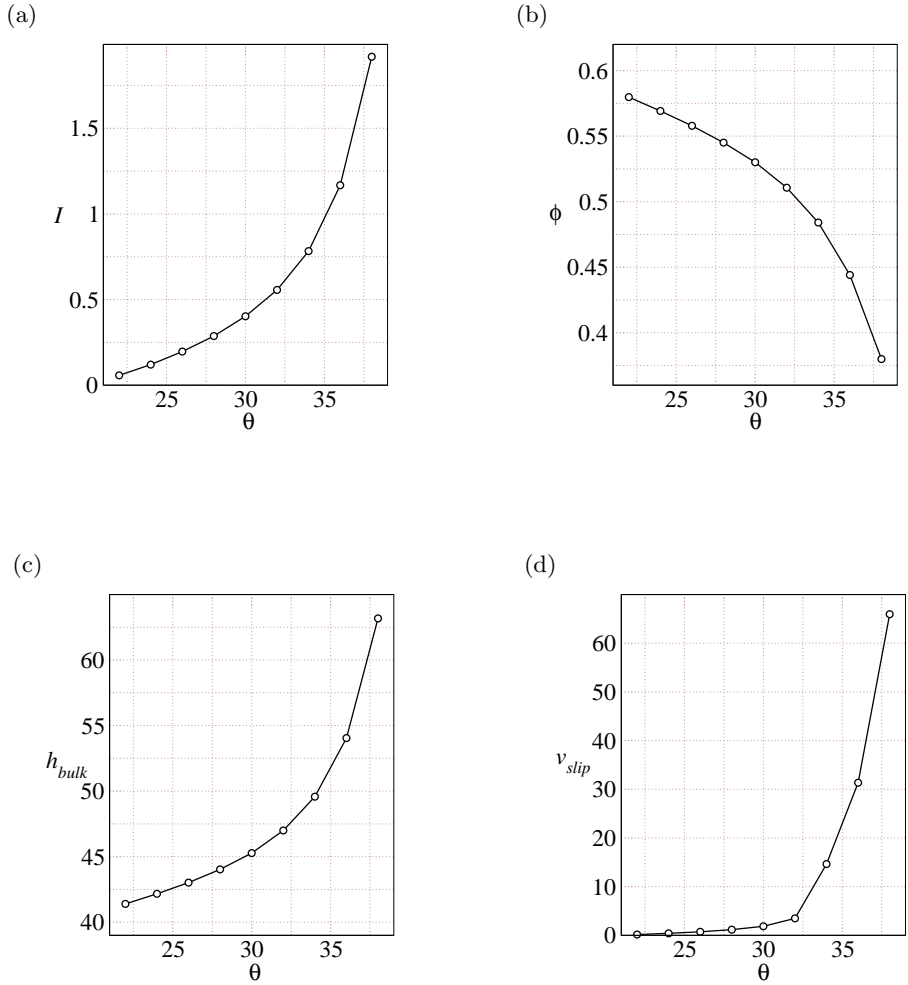

	\centering
	\includegraphics[scale=0.35]{Fig_13a.eps}
		\put (-145,155){(a)}
		\hspace{2cm}
	\includegraphics[scale=0.35]{Fig_13b.eps}
		\put (-145,155){(b)}
		 \hspace{2cm}
	
 \vspace{1.6cm}

 \includegraphics[scale=0.35]{Fig_13c.eps}
		\put (-145,155){(c)}
		\hspace{2cm}
	     \includegraphics[scale=0.35]{Fig_13d.eps}
		\put (-145,155){(d)}
		\hspace{2cm}
\caption {Variation of (a) inertial number $I$, (b) solids fraction $\phi$, (c) bulk layer height $H_{bulk}$ and (d) slip velocity $v_{slip}$ with inclination angle $\theta$ for restitution coefficient $e_n=0.5$.}
	\label{Figure_13}
\end{figure*}

Figs.~\ref{Figure_13}(a)-\ref{Figure_13}(d) summarise the results over the entire range of inclinations investigated in this study and show the variation of the inertial number $I_{bulk}$, solids fraction $\phi_{bulk}$, height of the bulk region of the flowing layer $h_{bulk}$ and the slip velocity at the base $v_{slip}$ as a function of inclination angle $\theta$ for restitution coefficient $e_n=0.5$. The slope of the curve for the lower angles differs significantly from that for the higher angles. This slope contrast is evident in the case of slip velocity shown in Fig.~\ref{Figure_13} for different flow regimes: dense flows with negligible slip velocity at lower angles and rapid, dilute flows with large slip velocities at higher angles. The steady-state solids fraction $\phi$ is found to decrease with the increase in the inclination angle $\theta$ and can be seen in Fig.~\ref{Figure_13}(b). The decrease in the solids fraction is followed by the dilation of the bulk flowing layer thickness $h_{bulk}=h_{initial}\times\phi_{max}/\phi_{steady}$ which increases with the inclination angle. This increase in the bulk flowing layer thickness with the inclination angle is shown in Figs.~\ref{Figure_13}(c). The slip velocity at the base of the chute is shown in Figs.~\ref{Figure_13}(d) which increases with the increase in inclination angle and changes its slope gradually beyond inclination angle $\theta=30^\circ$.

\subsection{Role of normal stress difference}

\begin{figure*}[hbtp]
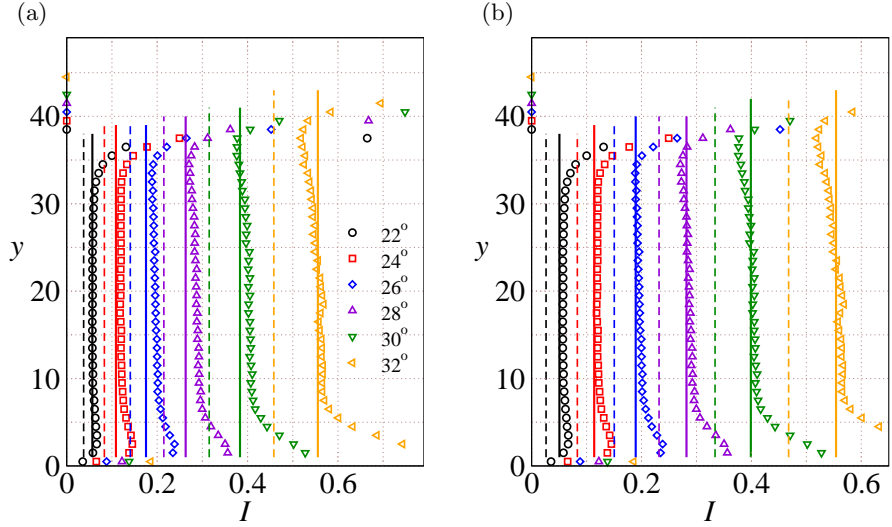

	\centering
    \includegraphics[scale=0.44]{Fig_14a.eps}\put (-155,190){(a)}
            \hspace{0.5cm}  
		\includegraphics[scale=0.44]{Fig_14b.eps}
		 \put (-155,190){(b)}
    
	\caption {Theoretical predictions of the inertial numbers using the a) JFP model, and b) modified rheological model with and without the effect of normal stress differences and its comparison with the DEM simulation results for inclination angles $\theta=22^\circ$ to $\theta=32^\circ$. Symbols represent the DEM simulation results. The dashed lines correspond to the theoretical predictions of the inertial number $I$ in which the normal stress differences have been ignored. The solid lines correspond to the theoretical predictions of the inertial number $I$ in which the normal stress differences have been considered.}
	\label{Figure_14}
\end{figure*}

In this section, we will discuss the effect of the first and the second normal stress differences on the flow properties of granular materials for the range of inclination angle $\theta=22^\circ$ to $\theta=32^\circ$. Figure \ref{Figure_14}(a) shows the theoretical predictions of the inertial number using the JFP model with and without considering the effect of the first and the second normal stress differences i.e. by assuming $f_1(I)=N_1/P=0$ and $f_2(I)=N_2/P=0$ in the latter case. Symbols represent the DEM simulation results. The dashed lines correspond to the theoretical predictions of the inertial number $I$ using the JFP model where the normal stress differences have been ignored whereas the solid lines correspond to the theoretical predictions of the inertial number $I$ using the JFP model where the normal stress differences have been considered. It is evident from figure \ref{Figure_14}(a) that the theoretical inertial number starts to deviate from the DEM simulation results as we increase the inclination angle $\theta$. Figure \ref{Figure_14}(b) shows the theoretical predictions of the inertial number using the modified rheological model with and without considering the effect of the first and the second normal state differences. The latter case assumes $f_1(I)=N_1/P=0$ and $f_2(I)=N_2/P=0$. As before, symbols represent the DEM simulation results. The dashed lines correspond to the theoretical predictions of the inertial number $I$ using the modified rheological model where the normal stress differences have been ignored whereas the solid lines correspond to the theoretical predictions of the inertial number $I$ using the modified rheological model where the normal stress differences have been considered. The deviation in the theoretical predictions of the inertial number $I$ from the DEM simulation results are observed in the case of the modified rheologial model as well and this deviation is more significant at higher inclinations. These results suggest that the role of the normal stress differences becomes crucial not only for large inclinations but also for small inclinations as well and one must consider these effects in order to accurately predict the flow properties. Neglecting the effect of the normal stress differences will lead to significant errors in the theoretical predictions.

\subsection{Oscillations in the steady flow}

\begin{figure*}[hbtp]
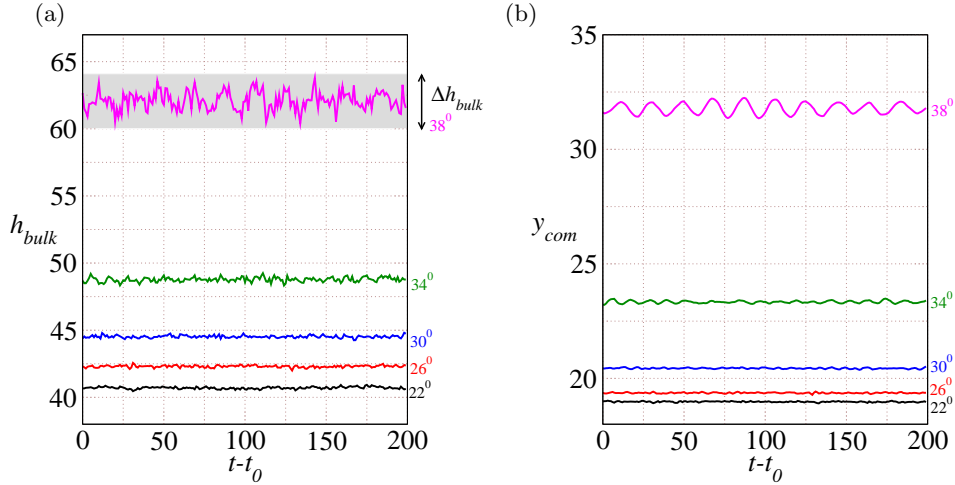

        \centering
 	\includegraphics[scale=0.4]{Fig_15a.eps}
 	\put (-170,175) {(a)}
 	\hspace{0.5cm}
	\includegraphics[scale=0.4]{Fig_15b.eps}
	\put (-170,175) {(b)}
	\caption{Steady state oscillating behavior of a) bulk flowing layer thickness $h_{bulk}$ 
    and b) center of mass $y_{com}$ at different inclinations for restitution coefficient $e_n=0.5$.}
\label{Figure_15}
\end{figure*}
\vspace{1.6cm}
\begin{figure*}[h!]
\centering
 	\includegraphics[scale=0.47]{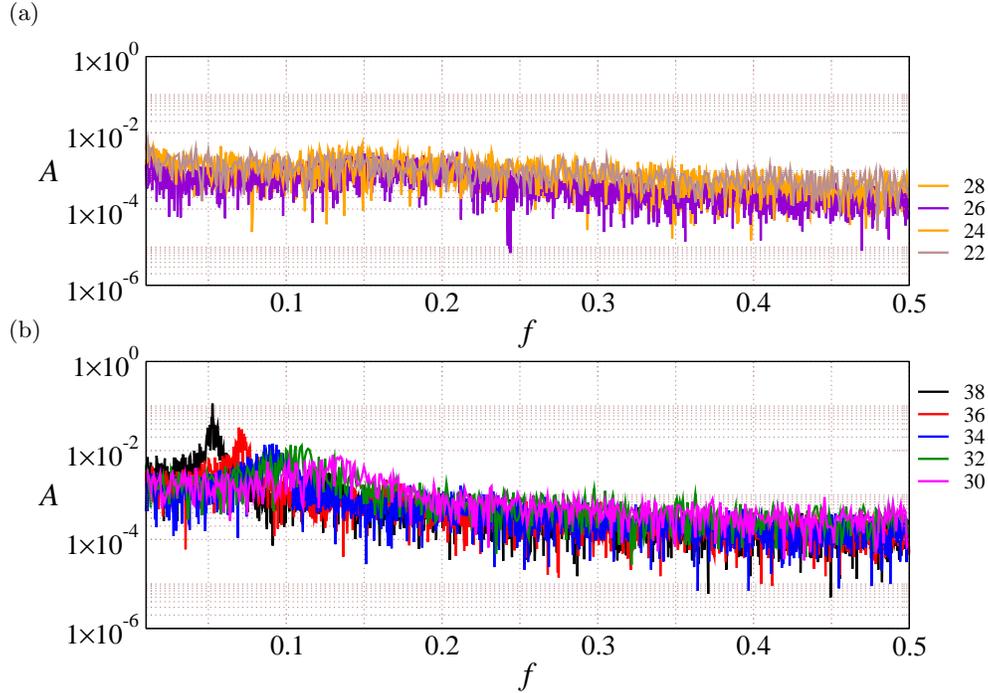}
  \put (-370,255) {(a)}
\put (-370,135) {(b)}
\caption{Fast Fourier transform analysis of the center of mass data for a) low to moderate inclination angles ranging from $\theta=22^\circ$ to $\theta=28^\circ$, b) moderate to large inclination angles ranging from $\theta=30^\circ$ to $\theta=38^\circ$.}
\label{Figure_16}
\end{figure*}

The steady-state oscillations in the bulk layer height $h_{bulk}$ are shown in Fig.~\ref{Figure_15}(a) over a time period of $200$ time units starting from a time instant $t_0$. 
Oscillations in the center of mass $y_{com}$ are shown in Fig.~\ref{Figure_15}(b). 
These height measurements are done after the flow has achieved steady kinetic energy (hence $t_0$ varies with the inclination angle $\theta$). 
 The oscillations in the bulk layer height $h_{bulk}$ as well as center of mass $y_{com}$ keep increasing with the inclination angle. While the difference between maximum and minimum bulk height is less than the particle diameter for inclinations $\theta\leq 30^\circ$, this variation becomes $\Delta h_{bulk}\approx 4d$ for $\theta= 38^\circ$ case (Fig.~\ref{Figure_15}(a)). These oscillations indicate that the role of density and height variations become important and cannot be ignored for granular flows at inertial numbers comparable to unity.

In order to investigate the time-periodic behavior of the layer at high inertial numbers, we performed a Fast Fourier Transform (FFT) analysis of the time series data of the center of mass position at steady state. 
Figure~\ref{Figure_16} shows the amplitude spectrum of the center of mass for different inclinations. 
The amplitude spectrum for inclinations $\theta=22^\circ$ to $\theta=28^\circ$ shows nearly uniform distribution for all frequencies as shown in Fig.~\ref{Figure_16}(a). At higher inclinations, a dominant frequency with a clear peak starts to appear and is shown in Fig.~\ref{Figure_16}(b). The occurrence of a dominant frequency in the amplitude spectrum confirms that the variation of layer height occurs with a characteristic time period at these large inclinations. The peak frequency keeps moving to lower values as the inclination angle increases, indicating that the time period of oscillations increases with the inclination angle. Figure~\ref{Figure_16}(b) also shows that the amplitude of oscillations at the $\theta=38^\circ$ is nearly an order of magnitude higher compared to those at $\theta=30^\circ$. 
\begin{figure*}[hbp]
\centering
 	\includegraphics[scale=0.47]{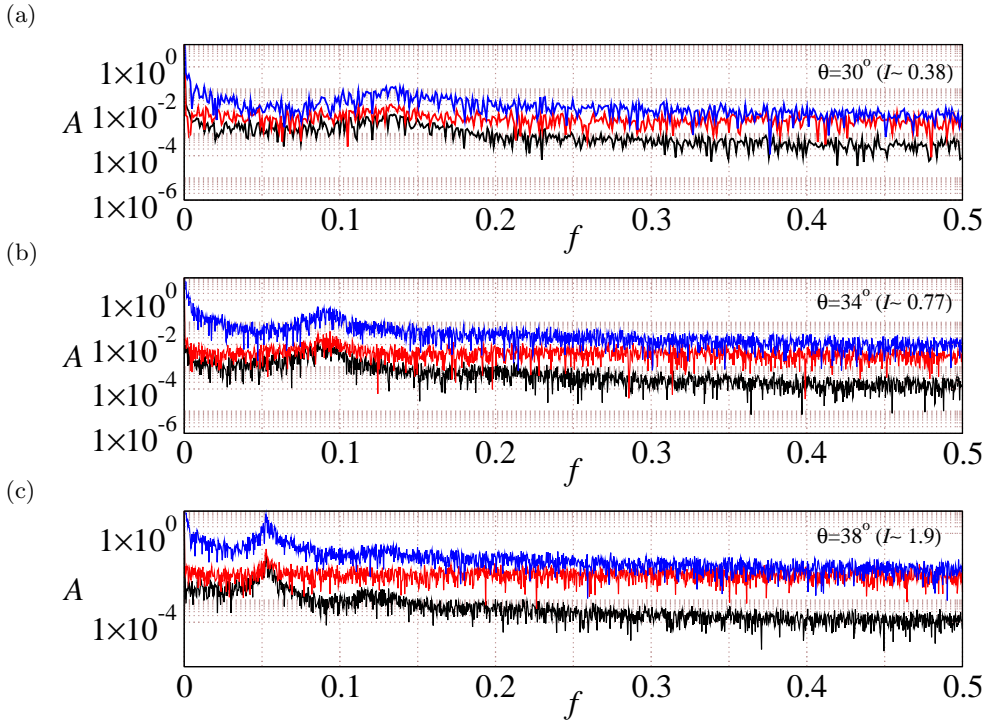}
 	\put (-370,265) {(a)}
 			\put (-370,175) {(b)}
 				\put (-370,85) {(c)}
    \caption{Fast Fourier transform analysis of the center of mass (shown in black line), bulk flowing layer thickness (shown in red line) and kinetic energy (shown in black line) for inclination angle a) $\theta=30^\circ$, b) $\theta=34^\circ$ and c) $\theta=38^\circ$. }
\label{Figure_17}
\end{figure*}

\begin{figure*}[hbp]
\centering
 	\includegraphics[scale=0.47]{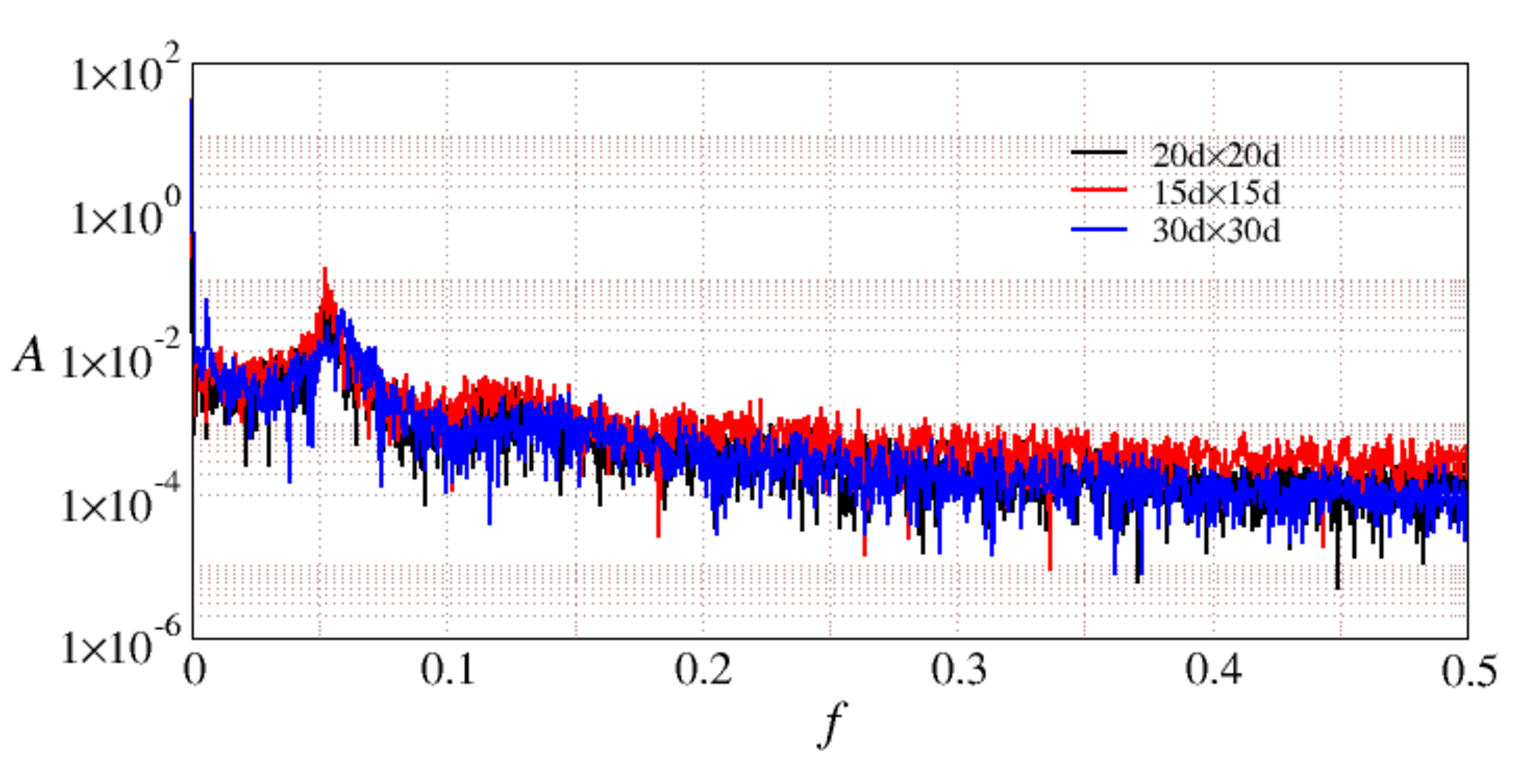}
 	\caption{\textcolor{black}{Fast Fourier transform analysis of the center of mass data at $\theta=38^\circ$ for three different simulation box sizes.} }
\label{Figure_18}
\end{figure*}

FFT analysis of the bulk layer height, center of mass, and kinetic energy data are shown for three different inclinations i.e. $\theta=30^\circ$, $\theta=34^\circ$, and $\theta=38^\circ$ as shown in Fig.~\ref{Figure_17}(a-c). The black solid line corresponds to the center of mass position, the red solid line corresponds to the bulk flowing layer thickness, and the black solid line corresponds to the kinetic energy. We find that the maximum amplitude of the spectrum is observed at the same frequency $f=0.05$ for different properties at the inclination $\theta=38^\circ$. For $\theta=34^\circ$ and $\theta=30^\circ$, the corresponding frequencies are $f=0.09$ and $f=0.13$ respectively. Hence we conclude that the oscillations at high inclinations affect these properties of the flow in a nearly identical fashion. 

\textcolor{black}{We also investigate the role of the simulation box size on the dominant frequency of oscillation and its corresponding amplitude. For this, we have considered three different simulation box size of base area $15d\times15d$, $20d\times20d$ and $30d\times30d$ where $d$ is the diameter of the . The initial height of the simulation setup at $t=0$ for all the three cases is same i.e. $40d$. Fig.~\ref{Figure_18} shows the FFT analysis of the center of mass data at $\theta=38^\circ$ for three different box sizes. We find that the maximum amplitude of the spectrum is observed at the same frequency $f=0.05$ at the inclination $\theta=38^\circ$ for three different simulation box size confirming that the amplitude and frequency of the oscillation is independent of the simulation box size.}

\section{Rheology at large inclinations}

\begin{figure*}[hbtp]
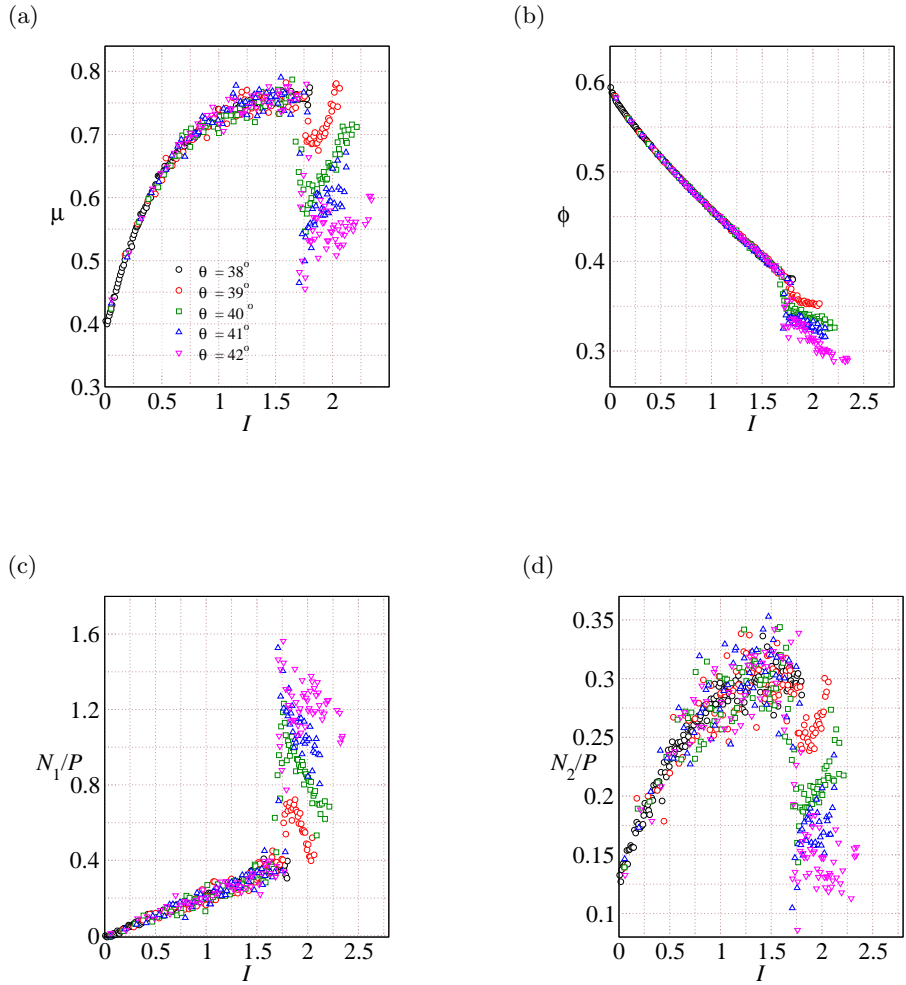

	\centering
	\includegraphics[scale=0.35]{Fig_19a.eps}
		\put (-145,155){(a)}
		\hspace{2cm}
	\includegraphics[scale=0.35]{Fig_19b.eps}
		\put (-145,155){(b)}
		 \hspace{2cm}
   
	\vspace{1.6cm}
 
 \includegraphics[scale=0.35]{Fig_19c.eps}
		\put (-145,155){(c)}
		\hspace{2cm}
	     \includegraphics[scale=0.35]{Fig_19d.eps}
		\put (-145,155){(d)}
		\hspace{2cm}
\caption {Variation of (a) Effective friction coefficient $\mu$, (b) solids fraction $\phi$, (c) first normal stress difference to the pressure ratio $N_1/P$ and (d) Second normal stress difference to the pressure ratio $N_2/P$ with the inertial number $I$ for inclination angles $\theta=38^\circ-42^\circ$ and for restitution coefficient $e_n=0.5$.}
	\label{Figure_19}
\end{figure*}

From the DEM simulations data obtained from the flowing layer thicknesses $L_y=40d$, we report the variation of the effective friction coefficient $\mu$, solids fraction $\phi$, the ratio of first normal stress difference to pressure $N_1/P$ and the ratio of second normal stress difference to pressure $N_2/P$ with inertial number $I$ for $e_n=0.5$ in Fig.~\ref{Figure_19}. The different coloured symbols represents the curve for different inclinations. The effective friction coefficient obtained from the simulations (shown using different symbols in Fig.~\ref{Figure_19}(a)) increases up to $I\approx1.8$ and shows a sudden decrease at large inertial numbers. There was a slight decrease in the effective friction coefficient $\mu$ at large inertial numbers $I$ reported at earlier sections upto the inclination angle $\theta=38^\circ$, however, the decrease was not prominent. The inclusion of $\mu-I$ data for inclination angles $\theta>38^\circ$ is shown to confirm the decrease in the effective friction coefficient at large inertial numbers.

\section{Conclusions}
We perform three-dimensional DEM simulations of slightly polydisperse, frictional, inelastic spheres flowing down a bumpy inclined plane varying over a wide range of inclination angles. We have considered two different flowing layer thicknesses: a thin layer with a settled layer height of $L_y=20d$ and a thick layer with a settled layer height of $L_y=40d$. We are able to observe steady flows for inclinations up to $\theta=38^\circ$ for $e_n=0.5$. The steady flow observed at $\theta=38^\circ$ corresponds to inertial number as high as $I\approx1.9$ and solids fraction as low as $\phi\approx0.38$. 
The flows at such high inertial numbers exhibit significant slip velocity at the bumpy base. However, they also exhibit a bulk core region with a nearly constant solids fraction as typically observed in case of dense flows. The solids fraction value in the bulk becomes as low as $\phi\approx0.38$ and the height of the layer increases up to $70d$ starting from a settled layer height of $40d$ for the highest inclination angle $\theta=38^\circ$ considered in this study.

Using the simulation data over a large range of inertial numbers, we find that key conclusions of the modified rheology previously reported for 2D systems in \citet{patro2021rheology} are observable in 3D as well. Given the range of inclination angles $\theta=22^\circ-38^\circ$ considered in this study, we were unable to observe the decreasing part of the $\mu-I$ curve at large inertial numbers. The inclusion of higher angle data $\theta>38^\circ$ may confirm the non-monotonic variation. However, even for this limited range of data, we find that the $\mu-I$ relation as per the modified rheological model does fit the data better compared to the JFP model. A power law relation describes the $\phi-I$ data well over the entire range of $I$. In addition, both the first and second normal stress difference laws, relating the normal stress differences to pressure ratio $f_1(I)=N_1/P$ and $f_2(I)=N_2/P$ with the inertial number $I$ is also proposed. We show that these normal stress difference laws needs to be considered to describe the rheology for large values of inertial numbers.
It is worth highlighting that the periodic oscillations in the flow properties at steady state become prominent at high inclinations. The bulk layer height, the center of mass position, and the kinetic energy show oscillations around the steady state value with a characteristic frequency beyond an inclination angle. The dominant frequency for all the properties at inclination $\theta=38^\circ$ shows a peak around $f=0.05$.
We conclude that the modified form of the $\mu-I$ rheology using the MK model along with the dilatancy law and both the normal stress difference law is able to predict the flow behavior for most of the bulk layer and is in good agreement with the simulation results up to inclination angle $\theta=38^\circ$.

\section*{ACKNOWLEDGEMENTS}
AT gratefully acknowledges the financial support provided by the Indian Institute of Technology Kanpur via the initiation grant IITK/CHE/20130338. 
\section*{DATA AVAILABILITY}
The data that support the findings of this study are available from the corresponding author upon request.

\section*{REFERENCES}
\nocite{*}
\bibliography{aipsamp}

\end{document}